\documentclass[english,aps, prl,superscriptaddress,twocolumn, showpacs, 10pt]{revtex4-1}
\usepackage[latin9]{inputenc}
\usepackage{amsmath}
\usepackage{amssymb}
\usepackage{graphicx}
\usepackage{esint}
\usepackage[T1]{fontenc}

\makeatletter
\@ifundefined{textcolor}{}
{%
 \definecolor{BLACK}{gray}{0}
 \definecolor{WHITE}{gray}{1}
 \definecolor{RED}{rgb}{1,0,0}
 \definecolor{GREEN}{rgb}{0,1,0}
 \definecolor{BLUE}{rgb}{0,0,1}
 \definecolor{CYAN}{cmyk}{1,0,0,0}
 \definecolor{MAGENTA}{cmyk}{0,1,0,0}
 \definecolor{YELLOW}{cmyk}{0,0,1,0}
 }

\usepackage{amsfonts}
\usepackage{epstopdf}
\usepackage{epsfig}
\usepackage{wrapfig}

\makeatother

\usepackage{babel}
\begin{document}

\title{Laser-induced cooling of broadband heat reservoirs}

\author{D. Gelbwaser-Klimovsky}
\affiliation{Weizmann Institute of Science, 76100
Rehovot, Israel}
\author{ K. Szczygielski}
\affiliation{Institute of Theoretical physics and Astrophysics,
University of Gda\'nsk, Poland}
\author{U. Vogl}
\altaffiliation{Present address: Max Planck Institute for the Science of Light, G\"unther-Scharowsky-Stra\ss e 1, Bau 24, 91058 Erlangen, Germany.}
\affiliation{Institut fur Angewandte Physik der Universit\"at Bonn, Wegelerstra\ss e 8, 53115 Bonn,
Germany}
\author{A. Sa\ss}
\affiliation{Institut fur Angewandte Physik der Universit\"at Bonn, Wegelerstra\ss e 8, 53115 Bonn,
Germany}
\author{R. Alicki}
\affiliation{Institute of Theoretical physics and Astrophysics,
University of Gda\'nsk, Poland}
\author{G. Kurizki}
\affiliation{Weizmann Institute of Science, 76100
Rehovot, Israel}
\author{M. Weitz}
\affiliation{Institut fur Angewandte Physik der Universit\"at Bonn, Wegelerstra\ss e 8, 53115 Bonn,
Germany}

\begin{abstract}
We explore, theoretically and experimentally, a method for cooling a broadband heat reservoir, via its laser-assisted collisions with two-level atoms followed by their fluorescence. This method is shown to be advantageous compared to existing laser-cooling methods  in  terms of its cooling efficiency, the lowest attainable temperature for broadband baths and   its versatility: it can cool down any heat reservoir, provided the laser is red-detuned from the atomic resonance. It is applicable to cooling down both dense gaseous and condensed media.
\end{abstract}

\pacs{37.10.De, 42.50.Wk, 34.90.+q, 05.90.+m}

\maketitle

\begin{figure}
\centering
\hspace{-1.2cm}
\includegraphics[width=0.45\textwidth]{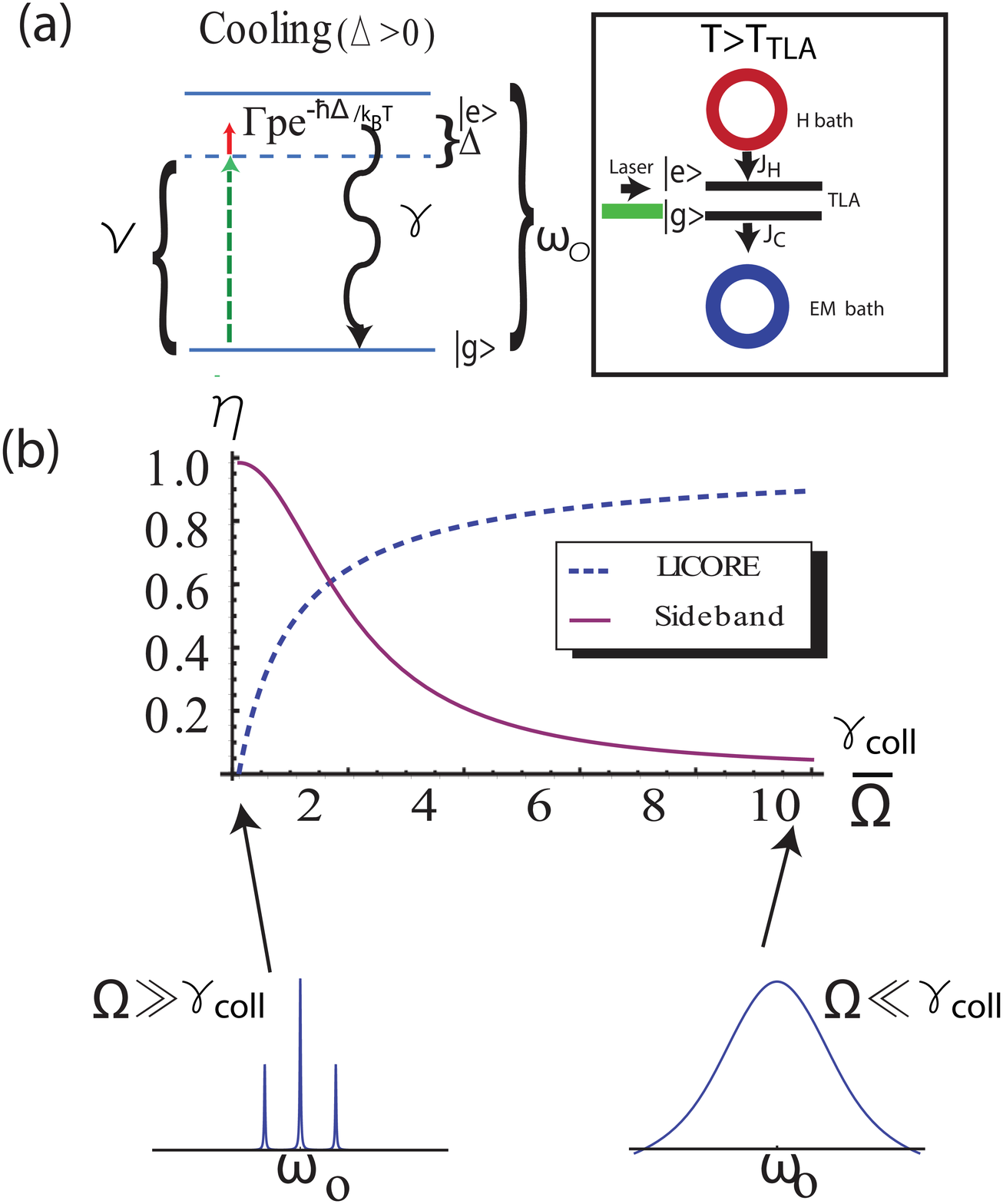}
\caption{(Color online) (a) Schematic  representation of cooling by  laser-induced collisional redistribution (LICORE). Left-- the level scheme along with the  corresponding laser detuning $\Delta$, collisional pumping rate $\Gamma_P$, and the atomic spontaneous decay rate $\gamma$. Right-- heat flow chart (TLA: two-level atom; EM: electromagnetic vacuum;  H bath: hot bath; C bath: cold bath). (b) Sideband cooling compared to LICORE. Top:  Efficiency as function of the band-width $\gamma_{coll}$ scaled by $\Omega$. While sideband cooling (solid red line) is highly efficient for resolved sidebands ($\Omega\gg\gamma_{coll}$), LICORE (dashed blue line) is much more efficient  for unresolved sidebands ($\Omega\ll\gamma_{coll}$). In sideband cooling $\Omega$ is the trapping frequency, while in LICORE $\Omega$ is the coupling Rabi frequency, taken to have the same value. The resolved sidebands spectrum  conforms to the Mollow triplet \cite{churmp98,sheiknatph07,schliessernatphys08}.}
\label{fig:currentsdirections(ULRICH)real}
\end{figure}

\textbf{Introduction} The advancement of quantum technologies  is demanding  new
cooling methods  \cite{taylorPRL11,RuschhauptJPB06,ErezNAT08,DomokosPRL02} as a follow-up on existing laser cooling
\cite{churmp98,sheiknatph07,schliessernatphys08,EpsteinNAT95,munganPRL97}.
A cooling method is assessed by its cooling power, thermodynamic efficiency (ratio of the cooling-power to the absorbed power), the minimal temperature it allows and its applicability under diverse  conditions.
Here we examine these thermodynamic characteristics for a simple method we have previously introduced \cite{voglnature09,VoglJMO11, VoglSpie12,SzczygielskiPRE13}, whereby heat is transferred from a reservoir by laser-assisted collisions to two-level atoms, and is subsequently emitted via  atomic fluorescence (Fig.\,\ref{fig:currentsdirections(ULRICH)real}a). This method, hereafter referred to as laser-induced collisional redistribution  (LICORE), is shown to be advantageous   compared to   sideband cooling \cite{schliessernatphys08,dehmeltnat76}, regarding the ability to cool down a \textit{broadband} heat reservoir. This ability is put to an experimental test for a hot bath of helium in collisional equilibrium with rubidium atoms and shown to be in  good agreement with theory.

The present work is essentially a thermodynamic analysis of the cooling experiment performed in \cite{VoglSpie12,voglnature09,VoglJMO11}. Its aim is to compare theory and experiment in LICORE setups and focus on its main thermodynamic implications.

\textbf{Theory}
The LICORE scheme consists of an ensemble of two-level atoms, driven by the laser, and permanently coupled  by \textit{elastic} collisions to a hot bath, and by spontaneous emission to a cold bath,  the electromagnetic vacuum, which is effectively at zero-temperature (Fig.\, \ref{fig:currentsdirections(ULRICH)real}a).

 For a hot bath at temperature $T$, the auto-correlation (coupling) spectrum, $G_H(\omega)=\int_{-\infty}^{\infty}e^{i \omega t} \langle B_{H}(t)B_{H}(0)\rangle dt$ satisfies the detailed-balance (Kubo-Martin-Schwinger, KMS) condition \cite{BreuerBOOK02}: $G_{H}(\omega)=e^{\frac{\hbar \omega}{k_B T}} G_{H}(-\omega),$ $B_H(t)$ being the  bath operator in the interaction picture and $k_B$ the Boltzmann constant. This detailed balance condition is akin to the Kennard-Stepanov ratio of absorption and emission rates \cite{kennardpr1918,stepanovspd57,Lakowiczbook,Moroshkinfut}. The spectral shape of $G_H(\omega)$ is immaterial to the occurrence of cooling,  but its peak and width affect the cooling rate, as shown below.

The steady-state solution of the master equation (see Appendix A) defines an \textit{effective temperature} $T_{\rm TLA}$, which is a measure of the stationary atomic  level-populations ratio $\rho_{ee}/\rho_{gg}$, where $\rho_{ee (gg)}$ is the excited- (ground-) state population of the two-level atom (TLA). The heat-flow direction between the  atoms and the hot bath is determined by $T_{\rm TLA}$. For large and positive  $\Delta$, i.e.\textit{ red detuning} from the atomic resonance ($\Delta \gg g$), its Boltzmann factor satisfies
\begin{equation}
e^{-\frac{\hbar \Delta}{k_{\rm B} T_{\rm TLA}}}\equiv\frac{\rho_{ee}}{\rho_{gg}}=
\frac{\Gamma_p e^{-\frac{\hbar \Delta}{k_{\rm B} T}}}{\Gamma_p +\gamma} \leq e^{-\frac{\hbar \Delta}{k_B T}},
\label{eq:ssc}
\end{equation}
\noindent where the collision-induced pumping rate  is
$ \Gamma_p= \left( \frac{2g}{\Delta}\right)^{2} G_{H}(|\Delta|).$ 
$g$ is the laser-atom resonant  coupling  and $\gamma$ is the spontaneous-emission rate. Hence, for $\Delta \gg g >0$, the  atoms are  effectively colder ($T_{\rm TLA} <
T$) than the hot bath. This is consistent with the  expressions obtained from our general theory \cite{SzczygielskiPRE13}  for  the heat current flowing from the hot bath to a two-level atom (see SI, Eq. (S20) for the exact expression)

\begin{equation}
J_{H}= \hbar {\Delta}\,\Gamma_p \,\frac{\gamma e^{-\frac{\hbar \Delta}{k_{\rm B} T}}}
{\gamma+\left(1+e^{-\frac{\hbar \Delta}{k_{\rm B} T}}\right)\Gamma_p} >0.
\label{current_app1}
\end{equation}

\begin{figure}
\centering
\includegraphics[width=0.65\textwidth]{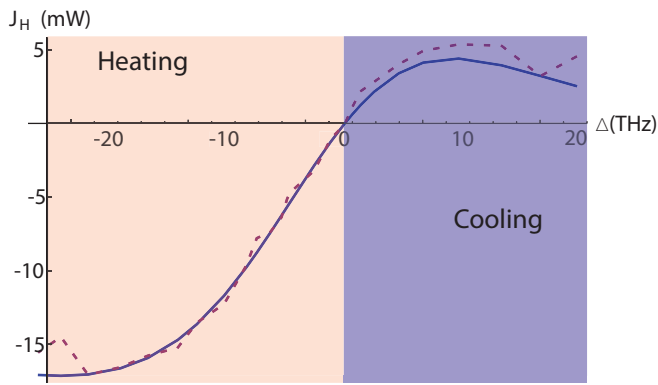}
\caption{(Color online)  Experimental results (red dashed line) compared to theoretical prediction (blue solid line) of the total cooling power as a function of the laser detuning $\Delta$. The  y-axis indicates the cooling power. For  red detuning, $J_H$ corresponds to  buffer gas cooling (purple, dark area), while blue detuning results in negative $J_H$, namely,  buffer gas heating (orange, light
area).   }
    \label{fig:withgaussinspect}
\end{figure}

\noindent    Consequently,  heat flows to the atoms, thereby  cooling the hot bath, provided $\Delta>0$,  for \textit{any} $T$, $\gamma$ and $\Gamma_p$.
The asymmetry between  cooling and heating  rates \textbf{of the buffer gas} in LICORE, as confirmed by the experimental data (Fig.\,\ref{fig:withgaussinspect}), is expressed by

\begin{equation}
\frac {-J_H(-\Delta)}{J_H(\Delta)} =  e^{\frac{\hbar \Delta}{k_{\rm B}T}}.
\label{current_rel}
\end{equation}

\noindent This asymmetry  is a consequence of two factors:
1) For  $|\Delta|\gg g$ the coupling strength to the cold bath (electromagnetic field)   is independent of the sign of $\Delta$.
2) The hot -bath (buffer gas) coupling to the TLA, $G_H(\Delta)$, satisfies the (KMS) detailed balance relation. The asymmetry  in \eqref{current_rel} conforms to the common intuition that it is easier to heat up than to cool down. LICORE can be understood as follows.  Collisions  broaden the atomic levels,  allowing the atom to absorb  a  photon  below resonance frequency ($\Delta=\omega_0-\nu>0$). Energy  flow from the hot bath   compensates for the energy mismatch between absorption at $\nu$ and  spontaneous emission  at $\omega_0$  to the electromagnetic vacuum  (cold bath).
The hot bath temperature is thereby reduced. In a perfectly isolated setup,  this  method would  cool the hot bath down  \textit{to an arbitrarily low temperature}. As the hot bath temperature goes down, the cooling power in \eqref{current_app1} is reduced but remains always positive.
 This is a salient advantage  compared to both Doppler cooling \cite{hanschoptcom75,winelandpra79} and sideband cooling  \cite{schliessernatphys08,EpsteinNAT95,munganPRL97,dehmeltnat76} that have fundamental \textit{minimal} cooling temperatures (see Table 1). The LICORE cooling efficiency  is a product of two factors, corresponding to a two-step process: First, a photon is absorbed by the atom with efficiency $\frac{P_{\rm abs}}{P_{\rm L}}$, $P_{\rm L}$ being the incident laser power. Next, the absorbed photon drives the cooling with the thermodynamic efficiency $\frac{J_H}{P_{\rm abs}}$.  Thus the total efficiency in the weak laser limit ($\Delta \gg g$, using \eqref{current_app1})

\begin{equation}
\eta = \frac{P_{\rm abs}}{P_{\rm L}} \frac{J_H}{P_{\rm abs}}=\Delta \frac{\Gamma_p}{P_{\rm L}}\frac{\gamma e^{-\frac{\hbar \Delta}{k_{\rm B}T}}}{\gamma+\left(1+e^{-\frac{\hbar \Delta}{k_{\rm B}T}}\right)\Gamma_p},
\label{eq:efficiency}
\end{equation}

 \noindent the absorbed power being (see SI)

\begin{equation}
P_{\rm abs}= \hbar \nu\,\Gamma_p \,\frac{\gamma e^{-\frac{\hbar \Delta}{k_{\rm B} T}}}
{\gamma+\left(1+e^{-\frac{\hbar \Delta}{k_{\rm B} T}}\right)\Gamma_p}.
\label{eq:pabs}
\end{equation}

\noindent The extracted energy from the hot bath follows the detuning,  and correspondingly  the cooling power increases, for moderate detuning in the linear regime (Fig. \ref{fig:withgaussinspect}). Yet, for large detuning laser absorption decreases, thereby  reducing the cooling power.
 The highest efficiency is achieved if all photons are absorbed, $(P_{\rm L}=P_{\rm abs}= \hbar \nu\Gamma_{p}\frac{\gamma}{\gamma+2\Gamma_{p}}$), with unity efficiency being reached if the energy shift from the redistribution becomes as large as the entire cooling light photon energy, $\hbar\nu$, yielding $J_H = P_{\rm abs}$. The efficiency of such a heat distributor can in principle be impressively high  (Fig. \ref{fig:currentsdirections(ULRICH)real}b). On the other hand, optical transitions in atoms do not allow for reasonable absorption rates at detunings of order of the absolute photon energy, and a more typical value for the high pressure buffer gas broadened ensemble is a detuning of order of the thermal energy, $\Delta \approx T$, see Fig. 2, corresponding to \(\Delta \approx 10\,\rm THz\) at \(500\,\rm K\) gas temperature. This yields a typical efficiency $J_H/ P_{\rm abs}\approx T/\omega_0$.

\begin {table*}[btp]
\begin{center}
\caption {Comparison of performance} \label{tab:title}
\begin{tabular}{|p{3.5cm}|c|c|c|}
\hline
 & $\Omega\ll\gamma$ & $\Omega\gg\gamma$ & Maximal thermodynamic efficiency $\left(\frac{J_{H}}{P_{\rm abs}}\right)$\tabularnewline
\hline
\hline
Doppler Cooling \cite{hanschoptcom75,winelandpra79} & $\frac{k_{\rm B}T_{\rm min}}{\hbar\gamma}\approx1/4$ & $\frac{k_{\rm B}T_{\rm min}}{\hbar\gamma}\approx1/4$ & $(m$ atom mass) $\frac{\hbar\nu}{2c^{2}m}\ll 1$\tabularnewline
\hline
&&&  Resolved bands $\frac{\Omega}{\nu}=\frac{\Omega}{\omega_{0}-\Omega} \lesssim 1$\\
Sideband cooling \cite{schliessernatphys08,EpsteinNAT95,munganPRL97,dehmeltnat76} &   $\frac{k_{\rm B}T_{\rm min}}{\hbar\Omega}\approx\frac{1}{\log\left(\frac{1+\gamma/4\Omega}{\gamma/4\Omega}\right)}\gg1$ & $\frac{k_{\rm B}T_{\rm min}}{\hbar\Omega}\approx\frac{1}{\log\left(\frac{1+\gamma^{2}/16\Omega^{2}}{\gamma^{2}/16\Omega^{2}}\right)}\ll1$ & \\
&&&Unresolved bands $\frac{\Omega}{\nu}=\frac{\Omega}{\omega_{0}-\Omega} \ll 1$\tabularnewline
\hline
Laser-induced collisional redistribution (LICORE) & $\frac{k_{\rm B}T_{\rm min}}{\hbar\Omega}\approx\frac{1}{4\log\left(\frac{\Delta}{g}\right)}\ll1$ & $\frac{k_{\rm B}T_{\rm min}}{\hbar\Omega}\approx\frac{1}{4\log\left(\frac{\Delta}{g}\right)}\ll1$ & Unresolved bands $\frac{\Omega}{\nu} \sim \frac{\Delta}{\nu}\lesssim 1$\tabularnewline
\hline
\end{tabular}
\par\end{center}
\end{table*}

To assess the merits of the LICORE method, we compare it (Table 1)  to existing laser cooling methods: (i) \textit{Doppler cooling}, which is based on the momentum change of an atom caused by the absorption and reemission of a photon \cite{sheiknatph07}; (ii) \textit{Sideband cooling} which takes advantage of the sidebands created by the interplay between internal and external degrees of freedom of  a species \cite{schliessernatphys08} to reduce its trapped-state occupation by  tuning a laser to the lower sideband. The latter method requires (Fig.\,\ref{fig:currentsdirections(ULRICH)real}b) the sidebands to be resolved, which amounts to  strong binding to the trap $\Omega \gg \gamma$, otherwise heating processes will compete with the cooling. We can draw an analogy between the trapping frequency $\Omega$ in sideband cooling and the coupling Rabi Frequency $\Omega=\sqrt{4g^2+\Delta^2}$ in LICORE since in both processes the modes to be cooled  have frequency $\Omega$. Yet, we note that in LICORE the suppression of  heating is achieved  for any ratio of $\gamma / \Omega$  provided there is  \textit{weak coupling} to the laser, $g\ll\Delta$, whence $\Omega \simeq \Delta\left(1+\frac{2g^2}{\Delta^2}\right)$, since the heating probability is  then proportional to $\frac{2g^{2}}{\Delta^{2}}$.

The  thermodynamic efficiency bound determines the fraction of the absorbed energy  used to cool down the bath. The Doppler cooling efficiency is low, because it
scales with the recoil parameter. Sideband cooling and LICORE cooling  may achieve similar efficiency  for  an \textit{ extremely narrowband} oscillator bath, the scenario foreseen by sideband cooling, but  LICORE may be highly efficient for a broadband bath, while sideband cooling efficiency will then be very low (Fig.\,\ref{fig:currentsdirections(ULRICH)real}b).

\textbf{Experiment}
 The cooling setup   of a high pressure cell with optical access contains a mixture of atomic rubidium vapor, with number density on the order of 10$^{16}$cm$^{-3}$, and  a noble buffer gas that serves as the hot bath, here helium gas with a number density of 1.5$\cdot$10$^{21}$cm$^{-3}$ to 6$\cdot$10$^{21}$cm$^{-3}$. The cell temperature is kept around 500\,K.

\begin{figure}
	\centering
		\includegraphics[width=0.45\textwidth]{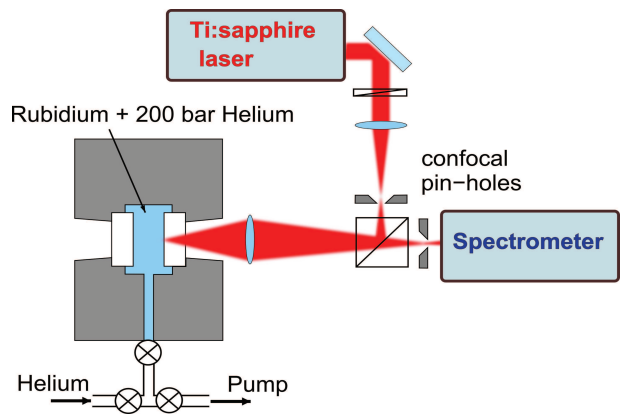}
	\caption{ Experimental setup for confocal spectroscopy on
high pressure helium-rubidium  gas mixtures. }
	\label{fig:Fig5laser}
\end{figure}

\begin{figure}
\center
\includegraphics[width=0.45\textwidth]{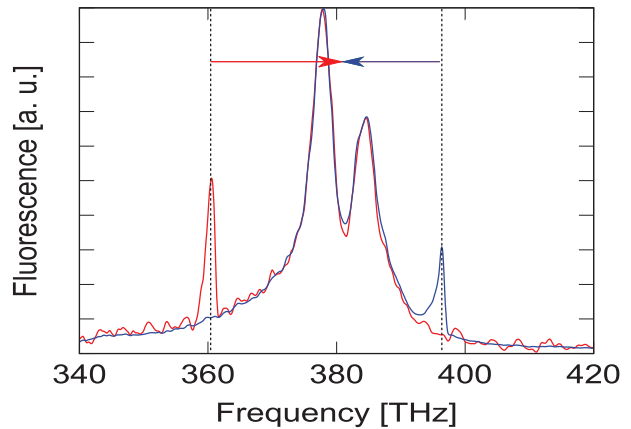}
\caption{ (Color online)  Fluorescence spectra for rubidium vapor at 530\,K with 200\,bar helium buffer gas for excitation frequency 360\,THz (red) and 401\,THz (blue). Arrows indicate the direction of the mean frequency shift of the scattered photons by the redistribution process towards the D-lines center. }
\label{fig:fluoryield}
\end{figure}

The confocal optical setup shown in Fig.\,\ref{fig:Fig5laser} is designed to characterize the spectral shift of the emitted fluorescence relative to the incident laser frequency. The collected atomic fluorescence radiation was analyzed
with an optical spectrometer. Typical experimental data for an excitation  laser frequency of 360\,THz (401\,THz), which is red (blue) detuned with respect to the D-lines resonances of the rubidium atoms, is shown in Fig.\, \ref{fig:fluoryield}.  The data displays typical pressure broadened fluorescence spectra, with the atomic rubidium D1- and D2-lines clearly visible, along with residual peaks of the exciting laser light. The figure clearly shows the redistribution of the fluorescence frequency to the center of the D-lines for both red and blue detuning respectively.

The cooling experiment is performed with a beam power of $P_{\rm L} =2.4$\,W on a 1\,mm beam diameter. On resonance, the measured optical density in the 1\,cm long gas cell, as derived from the transmitted laser power, reaches its maximum value of ca. $5$. A more detailed description of the experimental setup and the characterization of the system can be found in \cite{VoglJMO11}.

In the present analysis we use experimental data for helium \cite{VoglSpie12} instead of argon \cite{voglnature09} as a buffer gas, as the thermal properties of helium, i.e. thermal conductivity and heat capacity differ both by an order of magnitude from the respective values in argon. Our theoretical model shows for both cases (helium and argon) excellent agreement with the experimental data, which demonstrates that the physical mechanism of cooling here is largely independent of those quantities. Helium, due to its higher thermal conductivity may be better suited than other noble gases to couple the cooling scheme to a cooling load.

\textbf{Comparison of theory and experiment}
We have measured the center of the  fluorescence line  and $a(\nu)$, the photon absorption probability.
Upon identifying $P_{\rm abs}\equiv P_{\rm L} a({\nu}), \quad J_{H}=P_{L}a({\nu})\frac{\Delta}{\nu}$.
 We relate the theoretical cooling power, Eqs \eqref{current_app1} - \eqref{eq:pabs}, to its experimental counterpart. In our comparison, we account for laser absorption by many rubidium atoms in the cell.
	
The total cooling power can be calculated as: $ J_{H,tot}=\frac{N_{a}}{L} \int_0^L e^{- \alpha z} J_{H}(z)dz, $
 where  $N_{a}/L$ is the linear rubidium atomic density,  and the laser power is assumed to exponentially decay with z along the cell due to absorption. For a typical cooling laser frequency of 365\,THz, the measured total absorption in the cell is $67\%$, corresponding to an absorption length $l_{\rm abs} \equiv 1/\alpha \simeq9$\,mm.

A quantitative comparison  with theory requires an  ansatz for  the coupling (autocorrelation) spectrum to the buffer gas bath, the simplest being \textit{a constant function}:
$ G_{H}(\omega>0)=G_{H}(0).$
This ansatz means that the spectrum of collisions that couple the buffer gas  and the rubidium atoms is \textit{flat} within the absorption lineshape of the laser,  i.e. collisional effects are independent of $\Delta$. The factor $G_{H}(0)$  is determined from experimental data.
 We are aware that a more refined theory should account for the quasimolecular potential curves of rubidium-helium atom pairs  \cite{allardrmp82}.  Clearly,  heat leakage  (non-isolation) or molecule formation in the collisions  may stop the cooling. Experimentally, much lower temperatures are attainable provided that the cooling region is well isolated from heat leakage.

Figure \ref{fig:withgaussinspect}  compares  the theoretical (blue solid line) and measured (red dashed line) cooling power.  As predicted,  the buffer gas (hot bath) is cooled down or heated up  for red-detuned  or blue-detuned laser, respectively. The  difference between the measured and calculated curves reflects  the difficulty to precisely determine  the number of atoms  to which
the cooling power is proportional.
Nevertheless, the main experimental features, i.e. the cooling and heating range and their asymmetry, are in very good agreement with theory.

\textbf{Minimal temperature predictions}
In principle much lower temperatures  are attainable by LICORE than those observed by us, provided the technical issues discussed above are experimentally addressed. We again emphasize that the present theory does not account for quasimolecular character of the alkali-noble gas system.
LICORE cooling may reach lower temperatures than sideband or Doppler cooling, including  ultralow temperatures ($\frac{k_{\rm B} T_{\rm min}}{\hbar \Omega}\ll 1$) for both narrow and broadband spectra. Namely, $\Omega \simeq \Delta$, can be either narrow or broad compared to $\gamma$ and still yield very low temperatures, while sideband cooling can reach such temperature only for resolved sidebands (Table 1) $\Omega \gg \gamma$. Our theory predicts that for a given value of $\Omega$ at large detunings LICORE may reach lower temperature than Doppler cooling (Table 1).

In a perfectly isolated cell cooling stops only when $J_{H}=0$ at the minimum buffer gas temperature  obtainable from the \textit{exact} expression (SI)

\begin{equation}
e^{-\frac{\Omega}{T_{H}}}=\frac{\delta_{+}e^{-\frac{\nu_{+}}{T_{C}}}+\delta_{-}}{\delta_{+}+\delta_{-}e^{-\frac{\nu_{-}}{T_{C}}}}
\label{eq:premint}
\end{equation}

where the following short-hand notation is used:
$
\nu_{\pm}=\nu\pm\Omega,\quad\delta_{\pm}=\left(\frac{\Omega\pm\Delta}{2\Omega}\right)^{2}\frac{(\nu_{\pm})^{3}}{\omega_{0}^{3}}\gamma.
$ For a weak laser and positive detuning $(0<g\ll\Delta\ll \nu),$ $\nu_{+}\approx\omega_{o}$
and $1\gg\frac{\delta_{-}}{\delta_{+}}\gg e^{-\frac{\nu_{+}}{T_{C}}}$, the minimal temperature attains the value $\frac{k_{\rm B}T_{\rm min}}{\hbar\Omega}\approx\frac{1}{4\log\left(\frac{\Delta}{g}\right)}$ (see Table I).
However, the  LICORE cooling efficiency and rate are very low at such large detunings, as seen from  Eqs. (4) and (5) (Fig.\,1b).

\textbf{Discussion}
The main advantage of LICORE is that both the efficiency and the minimum temperature are \textit{adjustable by tunable laser parameters},  $\Delta$ and $g$, while for Doppler or sideband cooling they depend on  fixed parameters, spontaneous emission rate $\gamma$ and trapping frequency $\Omega$. Furthermore, LICORE cooling does not impose any restriction on either the hot- or cold- bath spectra, and is linear in the laser power. It thus  stands  in stark contrast to refrigeration  models \cite{Gelbwaserumach,KolarPRL12} where not only the modulation rate has to be above a critical rate,
but also non-overlapping (resolved) hot and cold bath spectra are required \cite{Gelbwaserumach}. However, the present scheme is a \textit{heat distributor and not a refrigerator}, as it cools down the hot bath (rather than the cold bath) at the expense of the laser power. Without the laser irradiation, the effect of collisions with the bath would only
be to decohere the  atoms,  rather than  change the bath  temperature.

Our combined experimental and theoretical results offer new fundamental and applied insights into the problem of laser-induced
cooling of a heat bath. The fundamental insight is that it is possible to run a laser-driven cooler that is not constrained by
the thermodynamic bounds governing a refrigerator or a sideband cooler, and therefore allows cooling a broadband bath down to much lower
temperature, and uniquely  high efficiency.
 The applied insight is that a new technological avenue may be opened for cooling diverse gaseous or condensed media by
laser-driven two-level atoms. Namely, the present method can in principle bypass the limitations of currently used  Doppler or sideband
cooling methods in terms of performance and versatility: the technique does not resort to \textit{auxiliary} atomic levels or to \textit{resolvable sidebands}; it
allows for \textit{arbitrary} (particularly, broadband) bath spectra and temperatures and only requires a red detuning of the cooling laser. \\

\textbf{Acknowledgments} We acknowledge the support of ISF and BSF
(G.K.), CONACYT (D.G.)  the DFG, grant no. We1748-15 (U.V., A.S., M.W.) and the Foundation
for Polish Science (TEAM project) co-financed by the EU (K.S. and
R.A.) and the University of Gdansk, grant no. 538-5400-B166-13 (K.S.).

%


%

\section*{Appendix}
\setcounter{equation}{0}
\renewcommand{\theequation}{A\arabic{equation}}

\subsection{A. Master Equation}

The total Hamiltonian is given by

\begin{equation}
H=H_S(t) + \sum_{j=H,C} [(H_{SB})_j+(H_{B})_j],\label{totH}
\end{equation}

\noindent where the $(H_{B})_J$ is the j-bath free Hamiltonian. Here the  laser-driven system  Hamiltonian is:

\begin{equation}
H_S(t)=\frac{\hbar}{2}\omega_{0}\sigma_{Z}+\hbar g\bigl(\sigma_{+}e^{-i\nu t}+\sigma_{-}e^{i\nu t}\bigr), \label{TLS}
\end{equation}
where  $\omega_{0}$ is
the (resonance) frequency of the two-level atom, $\sigma_Z$ and $\sigma_{\pm}=\sigma_X\pm i \sigma_Y$ are the appropriate spin-1/2 operators,  $\nu$ is the laser
frequency and g is the coupling strength between the laser and the
two-level atom. The laser detuning is $\Delta=\omega_0-\nu$.
The two-level atom coupling to the hot bath (H), via elastic collisions that do not change the two-level atom level populations, is described by

\begin{equation}
(H_{SB})_H=\hbar \sigma_{Z} \otimes B_H,
\label{eq:coupbg}
\end{equation}

\noindent where $B_H$ is the buffer gas bath operator. The coupling to the cold (C) bath (electromagnetic vacuum)  via spontaneous emission is given by

\begin{equation}
(H_{SB})_C= \hbar\sigma_{X} \otimes B_C,
\label{eq:}
\end{equation}

\noindent $B_C$ being the electromagnetic bath operator.

At steady-state, we may replace \eqref{TLS} of the main text by the
\emph{averaged Hamiltonian}

\begin{equation}
\bar{H}_S = \frac{\hbar}{2} \Delta \sigma_{Z} + \hbar g\sigma_{X},
\label{eq:aveH}
\end{equation}

\noindent where $\Delta=\omega_0-\nu$ expresses the laser detuning.
Then the coupling operator to the jth bath, $S_j$,  is decomposed into a Fourier  series
\begin{gather}
S_j(t) = U^{\dagger}(t) S_j U(t) = \notag \\
 \sum_{q= 0, \pm 1, \pm 2 ... \in \mathbf{Z}}\sum_{\{\bar{\omega}\}} S_{j,q}(\bar{\omega})e^{-i (\bar{\omega}+ q\nu)t}, \quad (j=H,C)
\label{fourier}
\end{gather}

The master equation has the following form with the Lindblad generator decomposed in a Floquet (harmonic) series
\begin{equation}
\dot{\rho}_S=\mathcal{L} \rho_S, \quad
\mathcal{L} = \sum_{j=H,C}\sum_{q= 0, \pm 1, \pm 2 ...}\sum_{\bar{\omega}} \mathcal{L}^{j}_{q\bar{\omega}},
\label{generator1}
\end{equation}

\begin{gather}
\mathcal{L}^j_{q \bar{\omega}}\rho = \notag \\ 
\frac{1}{2}\Big\{G_j(\bar{\omega}+ q\nu)\bigl([S_{j,q}(\bar{\omega})\rho, S^{\dagger}_{j,q}(\bar{\omega})] + [S_{j,q}(\bar{\omega}), \rho S^{\dagger}_{j,q}(\bar{\omega})]\bigr) + \notag \\
G_j(-\bar{\omega}- q\nu)\bigl([S^{\dagger}_{j,q}(\bar{\omega})\rho, S_{j,q}(\bar{\omega})] + [S^{\dagger}_{j,q}(\bar{\omega}), \rho S_{j,q}(\bar{\omega})]\bigr)\Big\}.
\label{generator_loc}
\end{gather}

Explicitly, in terms of
 $\Omega =\sqrt{4g^2+\Delta^2}$,  the laser-induced Rabi frequency, these coupling operators have the from

\begin{gather}
        S_C(\nu-\Omega) = \frac{\Delta - \Omega}{2 \Omega}\left(
\begin{array}{cc}
 0 & 1 \\
 0 & 0
\end{array}
\right), \qquad
        S_C(\nu) = \frac{g}{\Omega} \left(
\begin{array}{cc}
 1 & 0 \\
 0 & -1
\end{array}
\right), \qquad \notag\\
        S_C(\nu+\Omega) = \frac{\Delta + \Omega}{2 \Omega}\left(
\begin{array}{cc}
 0 & 0 \\
 1 & 0
\end{array}
\right).
    \end{gather}
    \label{Sigma1Freq}

\noindent By contrast
\begin{equation}
S_H(0) = \frac{\Delta}{\Omega}\left(
\begin{array}{cc}
 1 & 0 \\
 0 & -1
\end{array}
\right),
\qquad
S_H(\Omega) = \frac{2g}{\Omega}\left(
\begin{array}{cc}
 0 & 0 \\
 -1 & 0
\end{array}
\right).
\end{equation}

The heat currents flowing out of  the $j$-th  bath, which are the main observables of interest,
 obey the same harmonic (Floquet) expansion. They are given by
\begin{equation}
J_j(t) =\sum_{q=-1,0,1}\sum_{\bar{\omega}} J^j_{q\bar{\omega}}(t) .
\label{j-current}
\end{equation}

\begin{equation}
J^j_{q\bar{\omega}}(t)= - k_B T_j\mathrm{Tr}\bigl[( \mathcal{L}^j_{q\bar{\omega}}(t)\rho(t))\ln \tilde{\rho}^j_{q\bar{\omega}}(t)\bigr],
\label{curr_loc2}
\end{equation}
\begin{equation}
\tilde{\rho}^j_{q\bar{\omega}} = \frac{\exp\Bigl\{-\frac{(\bar{\omega}+q\nu)\bar{H}}{\bar{\omega}k_B T_j}\Bigr\}}{\mathrm{Tr}\exp\Bigl\{-\frac{(\bar{\omega}+q\nu)\bar{H}}{\bar{\omega} k_B T_j}\Bigr\}} .
\label{stat_loc1}
\end{equation}

The absorbed power is

\begin{equation}
P_{abs}=-J_C(t)-J_H(t)
\label{eq:pabssi}
\end{equation}

The population probabilities for the two-level atoms
satisfy the following Markovian master equation (in the interaction picture)
\begin{gather}
\frac{d\rho_{ee}}{dt}=
-(\Gamma_p+\gamma)\rho_{ee}+\Gamma_p e^{-\frac{\hbar |\Delta|}{k_B T}}\rho_{gg} \quad \Delta>0
\label{MEC}
\end{gather}

and for negative detuning
\begin{gather}
\frac{d\rho_{ee}}{dt}=
-\Gamma_p\rho_{ee}+(\Gamma_p e^{-\frac{\hbar |\Delta|}{k_B T}}+\gamma)\rho_{gg} \quad \Delta<0
\label{MEH}
\end{gather}

At steady-state Eq. \eqref{MEC} yields Eq. (1) of the main text.

\subsection{B. Heating regime}

It is instructive to consider, for comparison sake, the heating regime  obtained for $\Delta<0$.
In this regime $T_{TLA}$ is determined by
\begin{equation}
e^{-\frac{\hbar |\Delta|}{k_B T_{TLA}}}\equiv\frac{\rho_{ee}}{\rho_{gg}}=
\frac{\Gamma_p e^{-\frac{\hbar |\Delta|}{k_B T}}+\gamma}{\Gamma_p} \geq e^{-\frac{\hbar |\Delta|}{k_B T}}.
\label{eq:ssh}
\end{equation}
In this case the  atoms are hotter ($T_{TLA}>T$) than the hot bath, thus the heat flows from the  atoms, heating up the bath. In this case, the absorbed photons have more energy than the spontaneously emitted photons ($\omega_0>\nu$) and the extra energy is absorbed by the bath, heating it up. The heat current is then

\begin{equation}
J_H=- \hbar {|\Delta|}\,\Gamma_p \,\frac{\gamma}
{\gamma+\left(1+e^{-\frac{\hbar|\Delta|}{k_B T}}\right)\Gamma_p}<0 .
\label{current_app2}
\end{equation}

\noindent The physical process is similar to that in the cooling case except that the Boltzmann factor in (2) of the main text is here absent, causing an asymmetry of the cooling and heating rates;

\subsection{C. Minimum temperature}

The most general expression for the heat current flowing from the
hot bath is \cite{SzczygielskiPRE13}
\begin{gather}
J_{H}= \notag\\
N_{a}{\Omega}\,\Gamma_{p}\,\frac{e^{-\frac{\Omega}{T_{H}}}(\delta_{+}+\delta_{-}e^{-\frac{\nu_{-}}{T_{C}}})-(\delta_{+}e^{-\frac{\nu_{+}}{T_{C}}}+\delta_{-})}{\delta_{-}\left(1+e^{-\frac{\nu_{-}}{T_{C}}}\right)+\delta_{+}\left(1+e^{-\frac{\nu_{+}}{T_{C}}}\right)+\left(1+e^{-\frac{\Omega}{T_{H}}}\right)\Gamma_{p}}.\label{current_exact}
\end{gather}

 Under the conditions specified for Eq.  (6)

\begin{gather}
e^{-\frac{\Omega}{T_{H}}}\approx \notag \\
\frac{\delta_{-}}{\delta_{+}}=\left(\frac{\Omega-\Delta}{\Omega+\Delta}\right)^{2}\left(\frac{\nu_{-}}{\nu_{+}}\right)^{3}\approx\left(\frac{2\left(\frac{g}{\Delta}\right)^{2}}{1+\frac{\Omega}{\Delta}}\right)^{2}\left(\frac{\nu_{-}}{\nu_{+}}\right)^{3}\approx\left(\frac{g}{\Delta}\right)^{4}
\label{eq:mint}
\end{gather}

\subsection{D. Experimental setup}

The  setup is shown in the main text (Fig. 3). It is a high presure cell with optical access. It contains a mixture of  atomic rubidium vapor (whose density is of the order of 10$^{16}$cm$^{-3}$) with helium gas (whose density goes from 1.5$\cdot$10$^{21}$cm$^{-3}$ to 6$\cdot$10$^{21}$cm$^{-3}$). The cell walls are kept at around 500K. They are an extra heating source and its thermal contact with the buffer gas precludes the achievement of lower temperatures.
 Light from a titanium-sapphire laser is used to address the rubidium D1- and D2- resonances (near 377\,THz and 384\,THz, respectively).
Due to the high pressure of the  helium buffer gas,  the resonances are
 broadened to  linewidths of a few nanometer. Nevertheless, we may treat the rubidium atoms  as  laser-driven two-level atoms at the chosen (D-line)  transition.

\end{document}